\documentclass[a4paper]{article}
\usepackage{geometry}
\usepackage{amsmath}
\newcommand{\bv}{\begin{vmatrix}}
\newcommand{\ev}{\end{vmatrix}}

\newcommand{\bea}{\begin{eqnarray*}}
\newcommand{\eea}{\end{eqnarray*}}
\newcommand{\bean}{\begin{eqnarray}}
\newcommand{\eean}{\end{eqnarray}}

\date{ }
\title{$5$-dimensional Myers-Perry Black Holes Cannot be Over-spun by Gedanken Experiments}
\author{ Jincheng An$^1$\footnote{anjch@mail.bnu.edu.cn}, Jieru Shan$^1$\footnote{jierushan@mail.bnu.edu.cn}, Hongbao Zhang$^{1,2}$\footnote{hzhang@vub.ac.be}, Suting Zhao$^1$\footnote{stzhao@mail.bnu.edu.cn}   \\
\\
\\
$^1$Department of Physics, Beijing Normal University,
Beijing 100875, China\\
\\
$^2$Theoretische Natuurkunde, Vrije Universiteit Brussel, \\
and The International Solvay Institutes,
B-1050 Brussels, Belgium}

\begin{document}
\maketitle
\begin{abstract}
We apply the new version of gedanken experiment designed recently by Sorce and Wald, to over-spin the $5$-dimensional Myers-Perry black holes. As a result, the extremal black holes cannot be over-spun at the linear order. On the other hand, although the nearly extremal black holes could be over-spun at the linear order, this process is shown to be prohibited by the quadratic order correction. Thus no violation of the weak cosmic censorship conjecture occurs around the $5$-dimensional Myers-Perry black holes.
\end{abstract}

\section{Introduction}
When a singularity is not hidden behind a black hole horizon, such as to be seen by a distant observer, then it is called a naked singularity. The weak cosmic censorship conjecture (WCC) claims that naked singularity cannot be formed generically through gravitational collapse with physically reasonable matter\cite{penrose}. Even though there is still no general proof for this conjecture for the $4$-dimensional asymptotically flat spacetime, the supporting evidence has been accumulated and discussed for a few decades\cite{wald-cos}. Especially in $1974$, Wald suggested a gedanken experiment to test WCC by examining whether the black hole horizon can be destroyed by injecting a point particle\cite{wald72}. As a result, such a gedanken experiment turns out to be in favor of WCC.

However, there are two crucial assumptions underlying in the aforementioned gedanken experiment. First, the black hole in consideration is extremal in its initial state. Second, the analysis is performed only at linear order of the point particle's energy, angular moment, and charge. The violation of WCC occurs when one releases either of these two assumptions. In particular, as initiated by Hubeny in $1999$\cite{hubeny}, one can show that a nearly extremal Kerr-Newman black hole can be both over charged and over spun\cite{FY,hod,ted,third,SS}. In addition, when one takes into account the higher order terms in the energy, angular momentum, and charge of the test particle, an extremal Kerr-Newman black hole can even be destroyed\cite{gaozhang}. But nevertheless, these results may not indicate a true violation of WCC. Instead, in all of these situations, the test particle assumption may not be valid any more, so WCC may be restored when one carefully takes into consideration the self-force and finite-size effects\cite{BCK1,BCK2,ZVP,CB,CBSM}.

Motivated by this, Sorce and Wald have recently designed a new version of gedanken experiment\cite{wald2017}. Rather than analyzing the motion of the particle matter to obtain the condition for it to be absorbed by the black hole, they apply Wald formalism to completely general matter and obtain the first order variational inequality for the mass of the black hole by simply requiring the null energy condition on the horizon for the general matter, which reduces to that obtained in the old version of gedanken experiment when one regards the particle matter as the limiting case of the general matter\footnote{The similar first order variational inequality is also obtained in \cite{NQV}.}. Moreover, when the initial black hole is non-extremal, they also obtain a lower bound for the second order variation of the mass of the black hole, which somehow incorporates both the self-force and finite-size effects and can be used to prove that no violation of the Hubeny type can ever occur. This result further strengthens the belief in the validity of WCC in the $4$-dimensional asymptotically flat spacetime.

But nevertheless, the $4$-dimensional black holes have a lot of remarkable properties. It is natural to ask whether these properties are general features of black holes or whether they are unique to the world being $4$-dimensional. For example, neither the uniqueness theorem nor the spherical topology of the horizon persists for the black holes in higher dimensions\cite{Hawking,GS,LR}. Regarding WCC, fully non-linear numerical simulation has indicated that not only the $5$-dimensional black strings and black rings but also the $6$-dimensional Myers-Perry black holes can be destroyed by perturbations, with the horizons pinching into a generic formation of naked singularity\cite{LP,FKT,FKLT}. However, to the best of our knowledge, so far there has no numerical evidence for the similar formation of naked singularity by the perturbation of the 5-dimensional Myers-Perry black holes. This leads naturally to a restricted version of WCC in the $5$-dimensional asymptotically flat spacetime, namely the generic perturbations around the 5-dimensional Myers-Perry black holes give rise to no formation of naked singularity. Note that the gedanken experiment, no matter whether it is the new version or old one, does not appeal to the sophisticated full-blown numerical relativity, so it is rewarding to check the validity of our restricted version of WCC by gedanken experiments. Actually it has been shown in \cite{Portugal} that the $5$-dimensional Myers-Perry black holes can not be over-spun by the old gedanken experiment. But to obtain the above result analytically, not only does the scenario considered in \cite{Portugal} restrict into either singly rotating or equally rotating black holes but also focuses exclusively on the test particle falling in along the equator.  As alluded to before, compared to the old one, the new gedanken experiment does not require us to analyze the motion of bodies to determine what kind of trajectories will or will not be captured by the black hole horizon, so it is desirable to check the validity of such a WCC around the $5$-dimensional Myers-Perry black holes in a more general circumstance by performing such a new gedanken experiment. This is the purpose of the current paper. As a result, the $5$-dimensional general Myers-Perry black holes can not be over-spun by a generic matter perturbation, thus our restricted version of WCC holds in the $5$-dimensional asymptotically flat spacetime.


The structure of this paper is organized as follows. In Section \ref{variation}, we shall review the well-established Iyer-Wald formalism for any diffeomorphism covariant theory in any dimension, in particular, the first and second order variational identities.  In Section \ref{5D}, we restrict ourselves to the 5-dimensional Einstein theory and introduce the 5-dimensional Myers-Perry black holes. Here, taking into account that the relevant quantities for the $5$-dimensional Myers-Perry black holes are presented in the previous literature without an explicit derivation, we relegate such a derivation to Appendix A and B. In addition, we also rewrite these quantities in a convenient way for the later calculation. Then in Section \ref{Null}, we follow the idea in \cite{wald2017} to present the set-up for the new version of gedanken experiment, in particular, the first order perturbation inequality, as well as the second order perturbation inequality for the optimal first order perturbation of non-extremal black holes. With the above preparation, we conduct such a gedanken experiment to over-spin the extremal and nearly extremal 5-dimensional Myers-Perry black holes in Section \ref{ged}. We conclude our paper in the last section with some discussions.

\section{Iyer-Wald Formalism and Variational Identities}\label{variation}
Compared to the ordinary Lagrangian scalar $L$ constructed locally out of the metric $g_{ab}$ , its Riemann curvature,  and other matter fields $\psi$ as well as their symmetrized covariant derivatives, we prefer to start from a diffeomorphism covariant theory in an $n$-dimensional spacetime $M$ with a Lagrangian  $n$-form ${\mathbf L}=L{\mathbf \epsilon}_{a_1a_2...a_n}$, where ${\mathbf \epsilon}_{a_1a_2...a_n}$ is the canonical volume element associated with the metric $g_{ab}$\cite{wald94}. If we denote $\phi=(g_{ab},\psi)$ as all dynamical fields, then the variation of the Lagrangian gives rise to
\begin{eqnarray}\label{dL}
\delta \mathbf{L}=\mathbf{E} \delta\phi+d{\mathbf\Theta}(\phi,\delta\phi),
\end{eqnarray}
where the equations of motion read $\mathbf{E}=0$, and the $(n-1)$-form ${\mathbf\Theta}$ is called  the symplectic potential form. The symplectic current $(n-1)$-form is defined in terms of a second variation of ${\mathbf\Theta}$ as
\begin{eqnarray}
{\mathbf\omega}(\phi,\delta_1\phi,\delta_2\phi)=\delta_1{\mathbf\Theta}(\phi,\delta_2\phi)-\delta_2{\mathbf\Theta}(\phi,\delta_1\phi).
\end{eqnarray}
Associated with an arbitrary vector field $\chi^a$ be any smooth vector field on the spacetime $M$, one can further define a Noether current $(n-1)$-form as
\begin{eqnarray}\label{noether}
{\mathbf J}_\chi={\mathbf\Theta}(\phi,{\mathcal L}_\chi\phi)-\chi \cdot{\mathbf L}.
\end{eqnarray}
A straightforward calculation gives
\begin{eqnarray}
d{\mathbf J}_\chi=-{\mathbf E}{\mathcal L}_\chi\phi,
\end{eqnarray}
which indicates ${\mathbf J}_\chi$ is closed when the equations of motion are satisfied. Furthermore, it is shown in \cite{wald95} that the Noether current can always be expressed as \begin{eqnarray}\label{jqc}
{\mathbf J}_\chi=d{\mathbf Q}_\chi+\mathbf{C}_\chi,
\end{eqnarray}
where ${\mathbf Q}_\chi$ is called the Noether charge and $\mathbf{C}_\chi=\chi^a{\mathbf C}_a$ is called the constraint of the theory, which vanishes when the equations of motion are satisfied.

Now by keeping $\chi^a$ fixed and comparing the variations of (\ref{noether}) and (\ref{jqc}), we end up with
\begin{eqnarray}\label{fi}
d[\delta \mathbf{Q}_\chi-\chi\cdot{\mathbf\Theta}(\phi,\delta\phi)]={\mathbf\omega}(\phi,\delta\phi,\mathcal{L}_\chi\phi)-\chi\cdot\mathbf{E}\delta\phi-\delta \mathbf{C}_\chi.
\end{eqnarray}

In what follows, we shall focus exclusively on the case in which $\phi$ represents the exterior solution of a stationary black hole with $\xi^a$ the horizon Killing field
\begin{eqnarray}
\xi^a=(\frac{\partial}{\partial t})^a+\Omega^I(\frac{\partial}{\partial\varphi^I})^a
\end{eqnarray}
satisfying $\mathcal{L}_\xi\phi=0$, where $(\frac{\partial}{\partial\varphi^I})^a$ are Killing vector fields with closed orbits, and $\Omega^I$ are the corresponding angular velocities of the horizon.  Then the variation of (\ref{fi}) gives rise to
\begin{eqnarray}\label{si}
d[\delta^2 \mathbf{Q}_\xi-\xi\cdot\delta{\mathbf\Theta}(\phi,\delta\phi)]={\mathbf\omega}(\phi,\delta\phi,\mathcal{L}_\xi\delta\phi)-\xi\cdot\delta\mathbf{E}\delta\phi-\delta^2 \mathbf{C}_\xi.
\end{eqnarray}

Suppose that $\Sigma$ is a hypersurface surface with a cross section $B$ of the horizon and the spacial infinity as its boundaries, then it follows from (\ref{fi}) that
\begin{eqnarray}\label{FI}
\delta M-\Omega^I\delta J_I=\int_B[\delta \mathbf{Q}_\xi-\xi\cdot{\mathbf\Theta}(\phi,\delta\phi)]-\int_\Sigma\delta \mathbf{C}_\xi,
\end{eqnarray}
where we have resorted to the fact that the variation of the ADM conserved quantity $H_\chi$ conjugate to an asymptotic Killing vector field $\chi^a$ if it exists is given by
\begin{eqnarray}
\delta H_\chi=\int_\infty [\delta \mathbf{Q}_\chi-\chi\cdot{\mathbf\Theta}(\phi,\delta\phi)]
\end{eqnarray}
with $M$ the ADM mass conjugate to $(\frac{\partial}{\partial t})^a$ and $J_I$ the ADM angular momenta conjugate to $(\frac{\partial}{\partial\varphi^i})^a$. Similarly, it follows from (\ref{si}) that
\begin{eqnarray}\label{SI}
\delta^2 M-\Omega^I\delta^2 J_I=\int_B[\delta^2 \mathbf{Q}_\xi-\xi\cdot\delta{\mathbf\Theta}(\phi,\delta\phi)]-\int_\Sigma\xi\cdot\delta\mathbf{E}\delta\phi-\int_\Sigma\delta^2 \mathbf{C}_\xi+\mathcal{E}_\Sigma(\phi,\delta\phi),
\end{eqnarray}
where we have used the definition of the canonical energy of the perturbation $\delta\phi$ on $\Sigma$
\begin{eqnarray}
\mathcal{E}_\Sigma(\phi,\delta\phi)=\int_\Sigma\mathbf{\omega}(\phi,\delta\phi,\mathcal{L}_\xi\delta\phi).
\end{eqnarray}

\section{$5$-Dimensional Einstein Theory and Myers-Perry Black Holes}\label{5D}
For our purpose, we now specialize to the $5$-dimensional Einstein theory, i.e.,
\begin{eqnarray}
{\mathbf L}=\frac{1}{16\pi}R{\mathbf\epsilon}.
\end{eqnarray}
Whence we have
\begin{eqnarray}
\mathbf{E}^{ab}=-\frac{1}{16\pi}G^{ab}\mathbf{\epsilon}
\end{eqnarray}
with $G^{ab}$ Einstein tensor, and the symplectic potential $4$-form
\begin{eqnarray}
{\mathbf\Theta}_{abcd}=\frac{1}{16\pi}{\mathbf\epsilon}_{eabcd}g^{ef}g^{hi}(\nabla_i\delta g_{fh}-\nabla_f\delta g_{hi}).
\end{eqnarray}
The corresponding symplectic current reads
\begin{eqnarray}
\mathbf{\omega}_{abcd}=\frac{1}{16\pi}{\mathbf\epsilon}_{eabcd}w^e,
\end{eqnarray}
where
\begin{eqnarray}
w^a=P^{abcdef}(\delta_2g_{bc}\nabla_d\delta_1g_{ef}-\delta_1g_{bc}\nabla_d\delta_2g_{ef})
\end{eqnarray}
with
\begin{eqnarray}
P^{abcdef}=g^{ae}g^{fb}g^{cd}-\frac{1}{2}g^{ad}g^{be}g^{fc}-\frac{1}{2}g^{ab}g^{cd}g^{ef}-\frac{1}{2}g^{bc}g^{ae}g^{fd}+\frac{1}{2}g^{bc}g^{ad}g^{ef}.
\end{eqnarray}
Taking $\mathcal{L}_\chi g_{ab}=\nabla_a\chi_b+\nabla_b\chi_a$ into consideration and by a straightforward calculation, we can further obtain the Noether current as
\begin{eqnarray}
({\mathbf J_\chi})_{ abcd}=\frac{1}{8\pi}{\bf \epsilon}_{eabcd}\nabla_f(\nabla^{[f}\chi^{e]})+\frac{1}{8\pi}{\bf\epsilon}_{eabcd}G^{ef}\chi_f.
\end{eqnarray}
By comparing it with \eqref{jqc}, one can readily identify the Noether charge
\begin{eqnarray}
({\mathbf Q_\chi})_{abc}=-\frac{1}{16\pi}{\bf\epsilon}_{abcde}\nabla^d\chi^e,
\end{eqnarray}
and
\begin{eqnarray}\label{ce}
({\mathbf C}_f)_{abcd}=\frac{1}{8\pi}{\mathbf\epsilon}_{eabcd}G^e{}_f.
\end{eqnarray}

As to a $5$-dimensional spacetime which is asymptotically flat in the sense that in a Lorentzian coordinate system $\{x\}$ of flat metric $\eta_{ab}$ the metric behaves as
\begin{eqnarray}
g_{\mu\nu}=\eta_{\mu\nu}+O(\frac{1}{r^2}), \quad \partial_\rho g_{\mu\nu}=O(\frac{1}{r^3})
\end{eqnarray}
near the spatial infinity, one can show there exists a $4$-form $\mathbf{B}$ such that the ADM mass is given by
\begin{eqnarray}\label{admm}
M=\int_\infty\mathbf{Q}_\frac{\partial}{\partial t}-\frac{\partial}{\partial t}\cdot\mathbf{B}=\frac{1}{16\pi}\int_\infty dS r^k\delta^{ij}(\partial_ih_{kj}-\partial_kh_{ij})=\frac{1}{16\pi}\int_\infty dS r^k(\partial_ih_k{}^i-\partial_kh),
\end{eqnarray}
where $r^a=(\frac{\partial}{\partial r})^a$ and $h_{ij}$ is the spatial metric with the index raised and the tensor traced both by the background Euclidean metric $\delta_{ij}$. On the other hand, it is easy to see that the ADM angular momentum is given simply by
\begin{eqnarray}\label{adma}
J_I=-\int_\infty\mathbf{Q}_\frac{\partial}{\partial\varphi^I}.
\end{eqnarray}

The higher dimensional generalization of asymptotically flat stationary Kerr black hole solution to the vacuum Einstein equation was first obtained by Myers and Perry\cite{1986}, and its $5$-dimensional version reads
\begin{eqnarray}\label{metric}
ds^2&=&-dt^2+\frac{\mu}{\Xi}\left(dt-a_1\sin^2{\theta}d\varphi^1-a_2\cos^2{\theta}d\varphi^2\right)^2+\frac{r^2\Xi}{\Pi-\mu r^2}dr^2\nonumber\\
&&+\Xi d\theta^2+(r^2+a_1^2)\sin^2{\theta}(d\varphi^1)^2+(r^2+a_2^2)\cos^2{\theta}(d\varphi^2)^2
\end{eqnarray}
with
\begin{eqnarray}
\Xi=r^2+a_1^2\cos^2{\theta}+a_2^2\sin^2{\theta}, \quad \Pi=(r^2+a_1^2)(r^2+a_2^2),
\end{eqnarray}
where $\varphi^I\in[0,2\pi]$ and $\theta\in[0,\frac{\pi}{2}]$. As shown in Appendix A, the parameters $\mu$ and $a_I$ are related to the ADM mass and angular momenta respectively as
\begin{eqnarray}
M=\frac{3\pi\mu}{8}, \quad J_I=\frac{\pi \mu a_I}{4}.
\end{eqnarray}
Without loss of generality, we shall constrain $a_I$ to be non-negative in later discussions.

The spacetime singularity is located at $\Xi=0$ as the squared Riemann tensor is given by
\begin{eqnarray}
R_{abcd}R^{abcd}=\frac{24\mu^2}{\Xi^6}\left(4r^2-3\Xi\right)\left(4r^2-\Xi\right).
\end{eqnarray}
While $\Pi-\mu r^2=0$ is simply the coordinate singularity, and its roots can be expressed as follows
\begin{eqnarray}
&&r=\pm\frac{\sqrt{\mu-(a_1+a_2)^2}\pm\sqrt{\mu-(a_1-a_2)^2}}{2},
\end{eqnarray}
which are real if and only if
\begin{eqnarray}
\mu\geq (a_1+a_2)^2,
\end{eqnarray}
where the largest root $r_H$ designates the black hole event horizon with the area
\begin{eqnarray}\label{area}
A=2\pi^2\mu r_H.
\end{eqnarray}
As calculated out in Appendix B, the corresponding angular velocity and surface gravity of the horizon are given by
\begin{eqnarray}
\Omega^I=\frac{a_I}{r_H^2+a_I^2}, \quad \kappa=\frac{2r_H^2+a_1^2+a_2^2-\mu}{\mu r_H}\end{eqnarray}
In particular,
\begin{eqnarray}\label{ext}
\mu =(a_1+a_2)^2
\end{eqnarray}
corresponds to the extremal Myers-Perry black holes\footnote{When one of angular momenta vanishes, then the horizon disappears and the resulting spacetime is a naked singularity\cite{ring}.}. On the other hand, for
\begin{eqnarray}
\mu< (a_1+a_2)^2,
\end{eqnarray}
the Myers-Perry metric describes a naked singularity.

For our later convenience, we would like to rewrite the condition for the existence of the horizon in terms of the ADM mass and angular momenta as
\begin{eqnarray}\label{exh}
32M^3-27\pi(J_1+J_2)^2\geq 0.
\end{eqnarray}
By the same token, the relevant quantities associated with the horizon can be expressed as
\begin{eqnarray}
A=\frac{4\sqrt{\pi}(\alpha+\beta)}{3\sqrt{3}}, \quad \Omega^I=\frac{72\pi MJ_I}{(\alpha+\beta)^2+108\pi J_I^2}, \quad \kappa=\frac{\sqrt{3\pi}\alpha\beta}{8M^2(\alpha+\beta)},
\end{eqnarray}
where
\begin{eqnarray}
\alpha=\sqrt{32M^3-27\pi(J_1+J_2)^2}, \quad
\beta=\sqrt{32M^3-27\pi(J_1-J_2)^2}=\sqrt{\alpha^2+108\pi J_1J_2}.
\end{eqnarray}
Obviously, $\alpha\rightarrow 0$ corresponds to the near extremal limit.

\section{Null Energy Condition and Perturbation Inequalities}\label{Null}
As the new gedanken experiment designed in \cite{wald2017}, the situation we plan to investigate is what happens to the above Myers-Perry black holes when they are perturbed by a one-parameter family of the matter source according to Einstein equation
\begin{eqnarray}
G_{ab}(\lambda)=8\pi T_{ab}(\lambda)
\end{eqnarray}
around $\lambda=0$ with $T_{ab}(0)=0$. Without loss of generality but for simplicity, we shall assume all the matter goes into the black hole through a finite portion of the future horizon. With this in mind, we can always choose a hypersurface $\Sigma=\mathcal{H}\cup\Sigma_1$ such that it starts from the very early cross section of the unperturbed horizon $B_1$ where the perturbation vanishes, continues up the horizon through the portion $\mathcal{H}$ till the very late cross section $B_2$ where the matter source vanishes, then becomes spacelike as $\Sigma_1$ to approach the spatial infinity. In addition, we would like to work with the Gaussian null coordinates near the unperturbed horizon as
\begin{eqnarray}
g_{ab}(\lambda)=2(du)_{(a}[(dv)_{b)}-v^2\rho(\lambda)(du)_{b)}+v\pi_{b)}(\lambda)]+q_{ab}(\lambda),
\end{eqnarray}
where $v=0$ denotes the location of the unperturbed horizon, $u$ is the affine parameter of future directed null geodesic generators of $r=0$ surface for any metric in the family, $\pi_a$ and $q_{ab}$ are orthogonal to $k^a=(\frac{\partial}{\partial u})^a$ and $l^a=(\frac{\partial}{\partial v})^a$. As one can show, this choice of coordinates follows\cite{HW}
\begin{eqnarray}\label{crazy}
 \int_{B_1}\mathbf{Q}_\xi(\lambda)=\frac{\kappa}{8\pi} A_{B_1}(\lambda)
 \end{eqnarray}
if we further choose the bifurcate surface of the unperturbed horizon as $B_1$ in what follows when the black hole in consideration is non-extremal.

With the above preparation, (\ref{FI}) reduces to
\begin{eqnarray}
\delta M-\Omega^I\delta J_I=-\int_\Sigma\delta \mathbf{C}_\xi=-\int_\mathcal{H}\epsilon_{eabcd}\delta T^{ef}\xi_f=\int_\mathcal{H}\tilde{\mathbf{\epsilon}}\delta T_{ab}k^a\xi^b,
\end{eqnarray}
where $\tilde{\mathbf{\epsilon}}$ is the induced volume element on the horizon, satisfying $\epsilon_{eabcd}=-5k_{[e}\tilde{\epsilon}_{abcd]}$. Now if the null energy condition is satisfied such that $\delta T_{ab}k^ak^b|_\mathcal{H}\geq 0$, we have the first order perturbation inequality as
\begin{eqnarray}\label{FIE}
\delta M-\Omega^I\delta J_I\geq 0.
\end{eqnarray}
when the first order perturbation is optimal, namely saturates the above inequality, it obviously requires $\delta T_{ab}k^ak^b=0$\footnote{As pointed out in \cite{wald2017}, this optimal first order perturbation is achievable for instance by lowering the matter to the horizon. In particular, it is easy to see that the first order perturbation is always optimal for Klein-Gordon and Maxwell fields.}. Whence the first order perturbation of Raychaudhuri equation
\begin{eqnarray}
\frac{d\vartheta(\lambda)}{du}=-\frac{1}{3}\vartheta(\lambda)^2-\sigma_{ab}(\lambda)\sigma^{ab}(\lambda)-R_{ab}(\lambda)k^ak^b
\end{eqnarray}
tells us that $\delta\vartheta=0$ on the horizon if we choose a gauge in which the first order perturbed horizon coincides with the unperturbed one. Then it follows from (\ref{SI}) that

\begin{eqnarray}
\delta^2 M-\Omega^I\delta^2 J_I&=&-\int_\mathcal{H}\xi\cdot\delta\mathbf{E}\delta\phi-\int_\mathcal{H}\delta^2 \mathbf{C}_\xi+\mathcal{E}_\mathcal{H}(\phi,\delta\phi)+\mathcal{E}_{\Sigma_1}(\phi,\delta\phi)\nonumber\\
&=&\int_\mathcal{H}\tilde{\mathbf{\epsilon}}\delta^2 T_{ab}k^a\xi^b+\mathcal{E}_\mathcal{H}(\phi,\delta\phi)+\mathcal{E}_{\Sigma_1}(\phi,\delta\phi)\nonumber\\
&=&\int_\mathcal{H}\tilde{\mathbf{\epsilon}}\delta^2 T_{ab}k^a\xi^b+\frac{1}{4\pi}\int_\mathcal{H}\tilde{\mathbf{\epsilon}}\delta\sigma_{cd}\delta\sigma^{cd}\xi^a\nabla_au+\nonumber\\
&&\frac{1}{16\pi}(\int_{B_2}\hat{\mathbf{\epsilon}}\delta g^{cd}\delta\sigma_{cd}\xi^a\nabla_au-\int_{B_1}\hat{\mathbf{\epsilon}}\delta g^{cd}\delta\sigma_{cd}\xi^a\nabla_au)+\mathcal{E}_{\Sigma_1}(\phi,\delta\phi)\nonumber\\
&=&\int_\mathcal{H}\tilde{\mathbf{\epsilon}}\delta^2 T_{ab}k^a\xi^b+\frac{1}{4\pi}\int_\mathcal{H}\tilde{\mathbf{\epsilon}}\delta\sigma_{cd}\delta\sigma^{cd}\xi^a\nabla_au+\mathcal{E}_{\Sigma_1}(\phi,\delta\phi)\geq \mathcal{E}_{\Sigma_1}(\phi,\delta\phi).
\end{eqnarray}
Here $\hat{\mathbf{\epsilon}}_{abc}=k^d\tilde{\mathbf{\epsilon}}_{dabc}$ is the induced area volume on the cross section of the horizon. In addition, we have employed $k_e\delta g^{ef}|_\mathcal{H}=0$ in the second step, borrowed the result from \cite{HW} for $\mathcal{E}_\mathcal{H}(\phi,\delta\phi)$ in the third step, and used the reasonable assumption that our black hole is linearly stable in the fourth step\cite{DHS}, such that the first order perturbation will drive the system towards another Myers-Perry black hole at sufficiently late times, leading to the vanishing $\delta\sigma_{cd}$ at $B_2$. In the last step, we have again resorted to the null energy condition for the second order perturbation of matter source on the horizon. Now we are left out to calculate $ \mathcal{E}_{\Sigma_1}(\phi,\delta\phi)$. To achieve this,  we follow the trick invented in \cite{wald2017}, and write $\mathcal{E}_{\Sigma_1}(\phi,\delta\phi)=\mathcal{E}_{\Sigma_1}(\phi,\delta\phi^{MP})$, where $\delta\phi^{MP}$ is induced by the variation of a family of Myers-Perry black holes
\begin{eqnarray}
M^{MP}(\lambda)=M+\lambda\delta M, \quad J_I^{MP}(\lambda)=J_I+\lambda\delta J_I
\end{eqnarray}
with $\delta M$ and $\delta J_I$ chosen to be in agreement with the firs order variation of the above optimal perturbation by the matter source. Note that for this family, we have $\delta^2M=\delta^2J_I=\delta\mathbf{E}=\delta^2\mathbf{C}_\xi=\mathcal{E}_\mathcal{H}(\phi,\delta\phi^{MP})=0$. Thus applying (\ref{SI}) to this family, we have
\begin{eqnarray}
\mathcal{E}_{\Sigma_1}(\phi,\delta\phi^{MP})=-\int_{B_1}[\delta^2 \mathbf{Q}_\xi-\xi\cdot\delta{\mathbf\Theta}(\phi,\delta\phi^{MP})].
\end{eqnarray}
Note that $\xi^a=0$ at the bifurcation surface $B_1$ of a non-extremal black hole, thus we can further employ (\ref{crazy})  to obtain
\begin{eqnarray}
\mathcal{E}_{\Sigma_1}(\phi,\delta\phi^{MP})= -\frac{\kappa}{8\pi}\delta^2A_{B_1}^{MP},
\end{eqnarray}
where
\begin{eqnarray}
\delta^2A^{MP}=\frac{4\sqrt{\pi}}{3\sqrt{3}}(\frac{X}{\alpha^3}+\frac{Y}{\beta^3})
\end{eqnarray}
with
\begin{eqnarray}
&&X=96M\left\{27\pi M(J_1+J_2)(\delta J_1+\delta J_2)\delta M-9\pi M^2(\delta J_1+\delta J_2)^2+[8M^3-27\pi(J_1+J_2)^2](\delta M)^2\right\},\nonumber\\
&&Y=96M\left\{27\pi M(J_1-J_2)(\delta J_1-\delta J_2)\delta M-9\pi M^2(\delta J_1-\delta J_2)^2+[8M^3-27\pi(J_1-J_2)^2](\delta M)^2\right\}.\nonumber\\
\end{eqnarray}
Therefore we end up with our second order perturbation inequality
\begin{eqnarray}\label{SIE}
\delta^2 M-\Omega^I\delta^2 J_I\geq-\frac{\kappa}{8\pi}\delta^2A_{B_1}^{MP}=-\frac{1}{48M^2(\alpha+\beta)}(\frac{X\beta}{\alpha^2}+\frac{Y\alpha}{\beta^2}),
\end{eqnarray}
which, as demonstrated in \cite{wald2017}, has incorporated the self-force and finite-size effects.

\section{Gedanken Experiments to Over-spin a 5-Dimensional Myers-Perry Black Hole}\label{ged}
In this section, we will explore the gedanken experiments to over-spin both an extremal black hole and a nearly extremal black hole by the physical process described above.

 For an extremal black hole, the inequality (\ref{exh}) is saturated, i.e.,
\begin{eqnarray}\label{ext}
32M^3-27\pi(J_1+J_2)^2=0.
\end{eqnarray}
\eqref{exh} will be violated if we can perturb the black hole so that
\begin{eqnarray}\label{vio}
\delta M-\frac{9\pi(J_1+J_2)}{16M^2}(\delta J_1+\delta J_2)<0.
\end{eqnarray}
However, when the black hole is extremal, the angular velocity becomes
\begin{eqnarray}
\Omega^I=\Omega\equiv\frac{2 M}{3(J_1+J_2)}=\frac{9\pi(J_1+J_2)}{16M^2}.
\end{eqnarray}
Then our first order perturbation inequality tells us that \eqref{vio} cannot be satisfied, thus an extremal 5-dimensional Myers-Perry black hole cannot be over-spun by our gedanken experiment.

Now let us turn to the nearly extremal Myers-Perry black hole, which is characterized by the small $\alpha$ compared to $\sqrt{32M^3}$. To proceed, we define a function of $\lambda$ as
\begin{eqnarray}
f(\lambda)=32M(\lambda)^3-27\pi\left[J_1(\lambda)+J_2(\lambda)\right]^2,
\end{eqnarray}
for the aforementioned one-parameter family of perturbation by our gedanken experiment with $f(0)=\alpha^2$. If we can find an appropriate small value of $\lambda$ so that $f(\lambda)<0$, then our nearly extremal black hole will be over-spun. We shall assume the first order perturbation is optimal, i.e.,
\begin{eqnarray}
\delta M=\Omega^IJ_I=\Omega(\delta J_1+\delta J_2)-\frac{\alpha\Omega}{3\sqrt{3\pi}(J_1+J_2)}(\sqrt{\frac{J_2}{J_1}}\delta J_1+\sqrt{\frac{J_1}{J_2}}\delta J_2)+O(\alpha^2)
\end{eqnarray}
and expand $f(\lambda)$ to the quadratic order in both $\lambda$ and $\alpha$ as
\begin{eqnarray}
f(\lambda)=\alpha^2+\gamma_1\lambda+\gamma_2\lambda^2+O(\lambda^3, \lambda^2\alpha, \lambda\alpha^2,\alpha^3),
\end{eqnarray}
where
\begin{eqnarray}
\gamma_1&=&96M^2\delta M-54\pi(J_1+J_2)(\delta J_1+\delta J_2)=96M^2\delta M-\frac{81\pi(J_1+J_2)^2}{M}\Omega(\delta J_1+\delta J_2)\nonumber\\
&=&\frac{81\pi(J_1+J_2)^2}{M}[\delta M-\Omega(\delta J_1+\delta J_2)]+O(\alpha^2)=-6\sqrt{3\pi}(\sqrt{\frac{J_2}{J_1}}\delta J_1+\sqrt{\frac{J_1}{J_2}}\delta J_2)\alpha+O(\alpha^2),
\end{eqnarray}
and
\begin{eqnarray}
\gamma_2&=&48M^2[\delta^2 M-\Omega(\delta^2 J_1+\delta^2 J_2)]+96M(\delta M)^2-27\pi(\delta J_1+\delta J_2)^2+O(\alpha^2)\nonumber\\
&=&48M^2(\delta^2 M-\Omega^I\delta^2 J_I)+96M\Omega^2(\delta J_1+\delta J_2)^2-27\pi(\delta J_1+\delta J_2)^2+O(\alpha)\nonumber\\
&=&48M^2(\delta^2 M-\Omega^I\delta^2 J_I)+9\pi(\delta J_1+\delta J_2)^2+O(\alpha)\nonumber\\
&\geq&-\frac{48M^2\kappa}{8\pi}\delta^2 A_{B_1}^{MP}+9\pi(\delta J_1+\delta J_2)^2+O(\alpha)\nonumber\\
&=&-\frac{X}{\alpha^2}+9\pi(\delta J_1+\delta J_2)^2+O(\alpha)=27\pi (\sqrt{\frac{J_2}{J_1}}\delta J_1+\sqrt{\frac{J_1}{J_2}}\delta J_2)^2+O(\alpha).
\end{eqnarray}
If the $O(\lambda^2)$ term is ignored, then it is not hard to see that it is possible to make $f(\lambda)<0$ such that our black hole can be over-spun. However, if we take into account the $O(\lambda^2)$ term, then miraculously we have
\begin{eqnarray}
f(\lambda)\geq [\alpha-3\sqrt{3\pi}(\sqrt{\frac{J_2}{J_1}}\delta J_1+\sqrt{\frac{J_1}{J_2}}\delta J_2)\lambda]^2+O(\lambda^3, \lambda^2\alpha, \lambda\alpha^2,\alpha^3)
\end{eqnarray}
for the optimal first order perturbation. Thus we can conclude that when the second order correction is taken into consideration, a nearly extremal 5-dimensional Myers-Perry black hole cannot be over-spun either.

\section{Conclusion}\label{conclu}
We have performed the new version of gedanken experiment to check the restricted version of WCC in the $5$-dimensional asymptotically flat spacetime by trying to over-spin the $5$-dimensional Myers-Perry black holes. As a result, no violation of such a WCC is found at the linear order for an extremal $5$-dimensional Myers-Perry black hole. While for a nearly extremal $5$-dimensional Myers-Perry black hole, we find that the violation of Hubeny type occurs most dangerously under the optimal first order perturbation but our WCC is restored miraculously by the second order perturbation inequality. Our result indicates that the $5$-dimensional Myers-Perry black holes, once formed, will never be over-spun classically.



\section*{Acknowledgments}
J. An and S. Zhao are partially supported by the National Natural Science Foundation of China with Grant Nos.11375026, 11675015 and 11775022, as well as by ``the Fundamental Research Funds for the Central Universities" with Grant No.2015NT16. J. Shan is supported by the Beijing Research Fund for Talented Undergraduates.
H. Zhang is supported in part by the Belgian Federal Science Policy Office through the Interuniversity Attraction Pole P7/37, by FWO-Vlaanderen through the project G020714N, by the Vrije Universiteit Brussel through the Strategic Research Program ``High-Energy Physics". He is also an individual FWO Fellow supported by 12G3518N. In addition, H. Zhang would like to thank Jorge Sorce and Bob Wald for the numerous helpful clarifications about their work, as well as Pau Figueras for his helpful private communication on the stability of the $5$-dimensional Myers-Perry black holes and our restricted version of WCC in the $5$-dimensional asymptotically flat spacetime. Thanks are also due Stefan Hollands for his useful correspondence about the Gaussian null coordinates and related gauge choice.

\section*{Appendix A: ADM mass and angular momenta for $5$-dimensional Myers-Perry black holes}
In this appendix, we would like to calculate the ADM mass and angular momenta explicitly for our $5$-dimensional Meyers-Perry black holes.

First, in order to obtain the ADM mass in an efficient way, we can think of the ordinary derivative in the Lorentzian coordinate system as the covariant derivative compatible with the background flat metric, where the spatial derivative can also be regarded as the covariant derivative compatible with the background Euclidean metric. With this in mind, now we can proceed to calculate the ADM mass directly in the same coordinate system as that used in (\ref{metric}), where the Euclidean metric reads
\begin{eqnarray}
ds_E^2=dr^2+r^2d\theta^2+r^2\sin^2\theta (d\varphi^1)^2+r^2\cos^2\theta (d\varphi^2)^2.
\end{eqnarray}
The non-vanishing components of the corresponding Christoffel symbols can be obtained as follows
\begin{eqnarray}
&&\Gamma^r{}_{\theta\theta}=-r, \quad \Gamma^r{}_{\varphi^1\varphi^1}=-r\sin^2\theta, \quad \Gamma^r{}_{\varphi^2\varphi^2}=-r\cos^2\theta, \quad \Gamma^\theta{}_{r\theta}=\Gamma^{\varphi^1}{}_{r\varphi^1}=\Gamma^{\varphi^2}{}_{r\varphi^2}=\frac{1}{r},\nonumber\\
&&\Gamma^\theta{}_{\varphi^1\varphi^1}=-\Gamma^\theta{}_{\varphi^2\varphi^2}=-\sin\theta\cos\theta, \quad \Gamma^{\varphi^1}{}_{\theta\varphi^1}=\cot\theta, \quad \Gamma^{\varphi^2}{}_{\theta\varphi^2}=\tan\theta.
\end{eqnarray}
A straightforward calculation further gives
\begin{eqnarray}
r^k(\partial_ih_k{}^i-\partial_k h)=\frac{3\mu+2(a_1^2-a_2^2)\cos 2\theta}{r^3}+O(\frac{1}{r^4}).
\end{eqnarray}
Note that the $3$-sphere volume $dS=r^3\sin\theta\cos\theta d\theta d\varphi^1d\varphi^2$, then (\ref{admm}) follows the ADM mass
\begin{eqnarray}
M=\frac{3\pi\mu}{8}.
\end{eqnarray}
On the other hand, according to (\ref{adma}), one can write the ADM angular momenta as
\begin{eqnarray}
J_I=-\frac{1}{16\pi}\int_\infty dS (g^{tt}\partial_rg_{t\varphi^I}+g^{t\varphi^I}\partial_rg_{\varphi^I\varphi^I}),
\end{eqnarray}
By plugging the involved metric components into the above expression, we arrive at
\begin{eqnarray}
J_I=\frac{\pi\mu a_I}{4}.
\end{eqnarray}
\section*{Appendix B: Angular velocity and surface gravity of $5$-dimensional Myers-Perry black holes}
In this appendix, we shall provide an explicit calculation for the angular velocity and surface gravity of our $5$-dimensional Myers-Perry black hole horizon.

The strategy to calculate the angular velocity of the horizon is first to choose a new coordinate system
\begin{eqnarray}
\varphi^I=\varphi'^I+\omega^I(r,\theta)t
\end{eqnarray}
with the other coordinates unchanged such that the metric has no spatial-temporal cross component. This gives us a pair of algebraic equations for $\omega^I$ as
\begin{eqnarray}
g_{t\varphi^I}+g_{\varphi^I\varphi^J}\omega^J=0.
\end{eqnarray}
Then the angular velocity of the horizon can be obtained by plugging the involved metric components into the above equations and solving $\omega^I$ at the horizon as
\begin{eqnarray}
\Omega^I=\omega^I|_H=-(g^{-1})^{\varphi^I\varphi^J}g_{t\varphi^J}|_H=\frac{a_I}{r_H^2+a_I^2}.
\end{eqnarray}
A few lines of algebra reveal
\begin{eqnarray}
&&\xi_a\xi^a=g_{ab}\xi^a\xi^b=g_{\varphi^I\varphi^J}(\omega^I-\Omega^I)(\omega^J-\Omega^J)+g_{tt}-g_{t\varphi^I}(g^{-1})^{\varphi^I\varphi^J}g_{t\varphi^J}\nonumber\\
&&=g_{\varphi^I\varphi^J}(\omega^I-\Omega^I)(\omega^J-\Omega^J)-\frac{\Xi(\Pi-\mu r^2)}{\Xi(\Pi-\mu r^2)+\Pi\mu}
\end{eqnarray}
which means that our horizon is a Killing horizon. The surface gravity of the horizon can be calculated by the following formula
\begin{eqnarray}\label{sg}
-2\kappa\xi_a=\nabla_a(\xi_b\xi^b)=-\frac{\Xi(\Pi'-2\mu r)}{\Pi\mu}(dr)_a
\end{eqnarray}
on the horizon.
To obtain the surface gravity,  we are obviously required to write down $\xi_a$ explicitly. But the coordinate system used in ($\ref{metric}$) is ill-defined on the horizon.  So we would like to choose an in-going coordinate system such that
\begin{eqnarray}
dt=dt'-\frac{\Pi}{\Pi-\mu r^2}dr, \quad d\varphi^1=d\varphi'^1-\frac{a_1\Pi}{(\Pi-\mu r^2)(r^2+a_1^2)}dr, \quad d\varphi^2=d\varphi'^2-\frac{a_2\Pi}{(\Pi-\mu r^2)(r^2+a_2^2)}dr.\end{eqnarray}
In this new coordinate system, the metric reads
\begin{eqnarray}
ds^2&=&-dt'^2+\frac{\mu}{\Xi}\left(dt'-a_1\sin^2{\theta}d\varphi'^1-a_2\cos^2{\theta}d\varphi'^2\right)^2+2(dt'-a_1\sin^2\theta d\varphi'^1-a_2\cos^2\theta d\varphi'^2)dr\nonumber\\
&&+\Xi d\theta^2+(r^2+a_1^2)\sin^2{\theta}(d\varphi'^1)^2+(r^2+a_2^2)\cos^2{\theta}(d\varphi'^2)^2,
\end{eqnarray}
which is well behaved on the horizon, and yields
\begin{eqnarray}
\xi_a=\frac{\Pi-\mu r^2}{\Pi}[-(dt')_a+a_1\sin^2\theta (d\varphi'^1)_a+a_2\cos^2\theta(d\varphi'^2)_a]+\frac{r^2\Xi}{\Pi}(dr)_a.
\end{eqnarray}
Plugging its value on the horizon into (\ref{sg}), we end up with
\begin{eqnarray}
\kappa=\frac{\Pi'-2\mu r}{2\mu r^2}|_H=\frac{2r_H^2+a_1^2+a_2^2-\mu}{\mu r_H}.
\end{eqnarray}

\end{document}